\documentclass{kapproc} 

\RequirePackage{graphicx}%
\RequirePackage{epsf}%
\input{psfig.sty}
\usepackage{amssymb}

\upperandlowercase
\let\footnote\savefootnote
\let\footnotetext\savefootnotetext 
\let\footnoterule\savefootnoterule 
\kluwerbib 

\def\mh{\rm [M/H]}
\def\te{T_{\rm eff}}
\def\msol{\,\mbox{M}_\odot}
\def\mv{\mbox{M}_{V}}
\def\mk{\mbox{M}_{K}}
\def\mj{\mbox{M}_{J}}
\def\mbol{\mbox{M}_{bol}}
\def\mvol{\mbox{M}_\odot\, \mbox{pc}^{-3}}
\def\mlvol{(\log\mbox{M}_\odot)^{-1}\, \mbox{pc}^{-3}}

\def\ga{\,\hbox{\hbox{$ > $}\kern -0.8em \lower 1.0ex\hbox{$\sim$}}\,}
\def\la{\,\hbox{\hbox{$ < $}\kern -0.8em \lower 1.0ex\hbox{$\sim$}}\,}

\def\wig#1{\mathrel{\hbox{\hbox to 0pt{%
\lower.5ex\hbox{$\sim$}\hss}\raise.4ex\hbox{$#1$}}}}

\def\apj{{\rm ApJ}, }
\def\aap{{\rm A\&A}, }
\def\aj{{\rm AJ}, }
\def\mnras{{\rm MNRAS}, }
\def\pasp{{\rm PASP}, }

\begin{document}


\articletitle{The initial mass function : from Salpeter 1955 to 2005}


\chaptitlerunninghead{The initial mass function in 2005}



 \author{Gilles Chabrier}
 \affil{Ecole Normale Sup\'erieure de Lyon, CRAL, Lyon, France}
 \email{chabrier@ens-lyon.fr}

\begin{abstract}
Fifty years after Ed Salpeter's seminal paper, tremendous progress both on the observational and theoretical
sides allow a fairly accurate determination of the Galactic IMF not only down to the
hydrogen-burning limit but into the brown dwarf domain. The present review includes the most recent observations of low-mass stars and brown dwarfs
to determine this IMF and the related Galactic mass budget. The IMF definitely exhibits a similar behaviour in various environments, disk,
young and globular clusters, spheroid.
Small scale dissipation of large scale compressible MHD turbulence seems to be the
underlying triggering mechanism for star formation.
Modern simulations of compressible MHD turbulence yield an IMF consistent with the one derived from observations.
\end{abstract}

\section{Introduction}
The determination of the stellar initial mass function (IMF) is one of the holly grails of
astrophysics. The IMF determines the baryonic
content, the chemical enrichment and the evolution of galaxies, and thus the universe's light and baryonic matter evolution. The IMF
provides also an essential diagnostic to understand the formation of stellar and substellar
objects. In this review, we outline the current determinations of the IMF in different
galactic environments, measuring the progress accomplished since \cite{Sal55} seminal paper
50 years ago. We also examine this IMF in the context of
modern theories of star formation. A more complete review can be found in \cite{Chabrier03a} but very recent results are included in the present paper.

\section{Mass-magnitude relations}

Apart from binaries of which the mass can be determined eventually by use of Kepler's third law, the
determination of the MF relies on the transformation of the observed luminosity function (LF), $\Phi=dN/dM$, i.e. the number of stars $N$ per absolute magnitude interval $dM$.
This transformation involves the derivative of a mass-luminosity relationship, for a given age $\tau$, or
preferentially of a mass-magnitude relationship (MMR),
${dn\over dm}(m)_\tau=({dn\over dM_\lambda(m)})\times ({dm\over dM_\lambda(m)})^{-1}_\tau$, which applies directly in the observed magnitude $M_\lambda$ and avoids the use of often ill-determined bolometric and $\te$-color corrections.

\begin{figure}[ht]
\sidebyside
{\centerline{\psfig{file=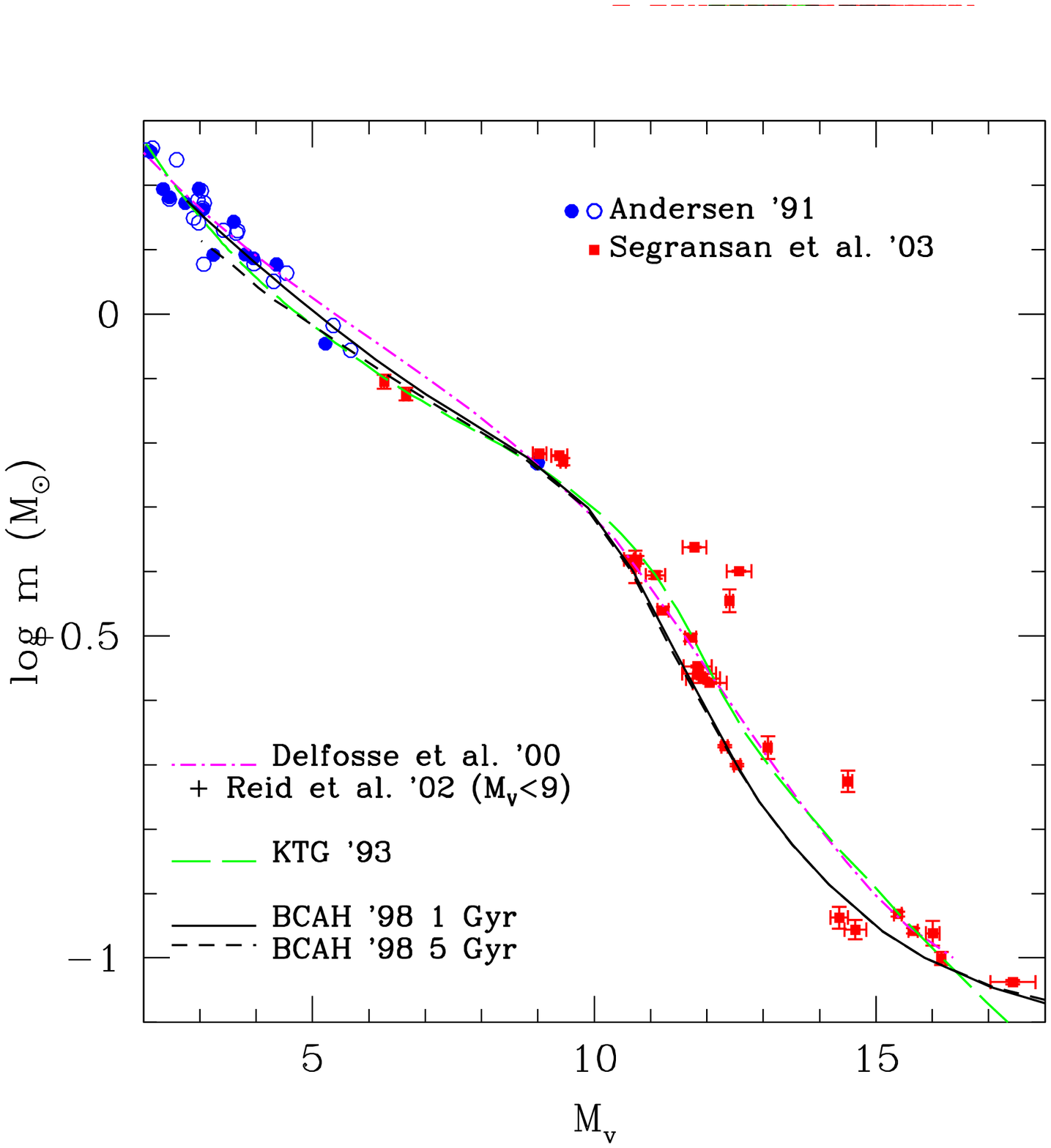,height=3.4in}}
\caption{Comparison of the observed and theoretical m-$\mv$ relation.}
\label{masMv}}
{\centerline{\psfig{file=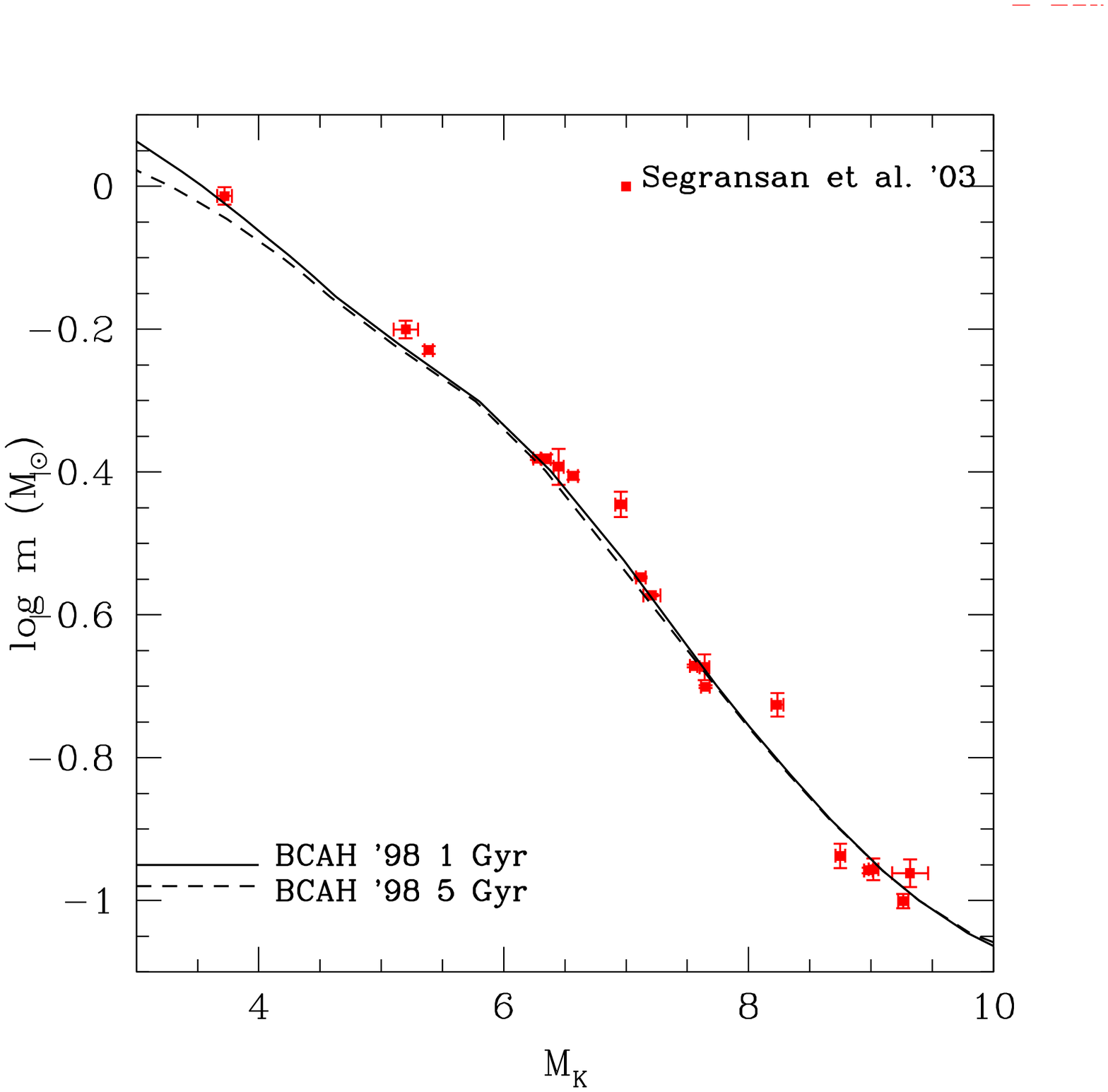,height=3.4in,width=2.8in}}
\caption{Same for the m-$\mk$ relation.}
\label{masMk}}
\end{figure}

Figure \ref{masMv} displays the comparison of the \cite{Andersen91} and \cite{Seg03} data in the V band with different theoretical MMRs, namely the parametrizations of \cite{KTG93} (KTG), \cite{Reid02} for $\mv <9$ complemented by \cite{Del00} above this limit and the models of \cite{BCAH} (BCAH) for two isochrones. The KTG MMR gives an excellent parametrization of the data
over the entire sample but fails to reproduce the flattening of the MMR near the low-mass end, which arises from the onset of degeneracy near the bottom of the main sequence (MS), yielding too steep a slope. The \cite{Del00} parametrization, by construction, reproduces the data in the $\mv$=9-17 range. For $\mv< 9$,
however, the parametrization of Reid et al. (2002) misses a few data, but more importantly does not yield the
correct magnitude of the Sun for its age. The BCAH models give an excellent representation
for $m\ga 0.4\msol$. Age effects due to stellar evolution start playing a role above $m\sim 0.8\,\msol$,
where the bulk of the data is best reproduced for an age 1 Gyr, which is consistent with a
stellar population belonging to the young disk ($h< 100$ pc).
Below $m\sim 0.4 \,\msol$, the BCAH MMR clearly differs from the \cite{Del00} one. Since we
know that the BCAH models overestimate the flux in the V-band, due to still incomplete molecular opacities,
we use the \cite{Del00} parametrization in this domain. The difference yields a maximum $\sim 16$\% discrepancy in the mass determination near $\mv \sim 13$.
Overall, the general agreement can be considered as very good, and the inferred error
in the derived MF is smaller than the observational error bars in the LF. The striking result is
the amazing agreement between the theoretical {\it predictions} and the data in the K-band
(Figure \ref{masMk}), a more appropriate band for low-mass star (LMS) detections.

\section{The disk and young cluster mass function}

\subsection{The disk mass function}

A V-band nearby LF $\Phi_V$ can be derived by combining Hipparcos parallax data (ESA 1997), which are essentially complete for $\mv < 12$ at $r_{comp}$=10 pc, and the sample of nearby stars 
with ground-based parallaxes for $\mv>12$ with $r_{comp}$=5.2 pc (\cite{Dahn86}).
Such a sample has been reconsidered recently by \cite{Reid02}, and extended to r=8 pc by \cite{Reid04}.
The revised sample agrees within $\sim$1$\sigma$ with the previous one,
except for $\mv \ga 14$, where the 2 LFs differ at the $\sim$2$\sigma$ limit.
The 5-pc LF was also obtained in the H and K bands by \cite{HMcC90}.  

\begin{figure}[ht]
\sidebyside
{\centerline{\psfig{file=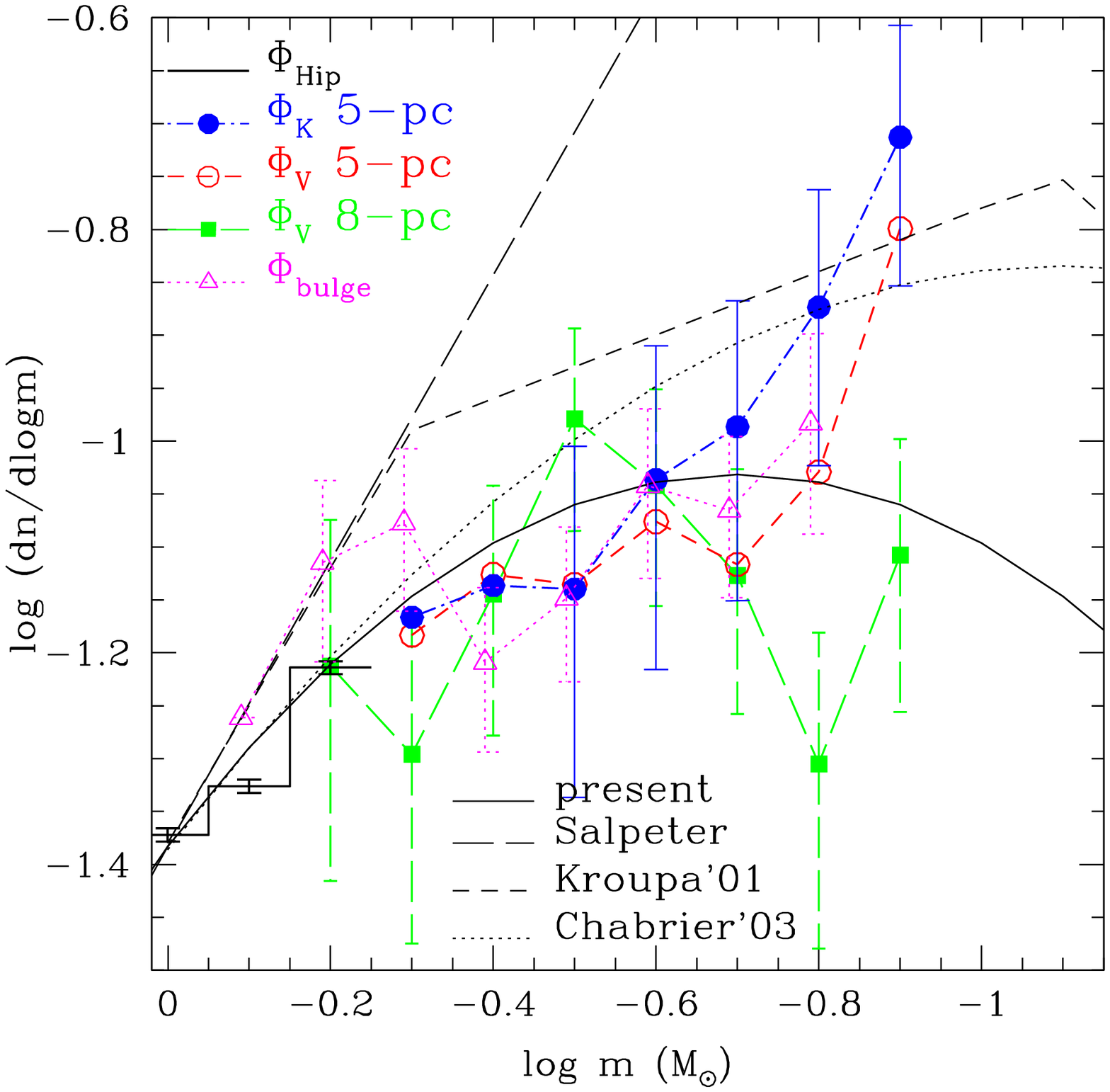,height=3in}}
\caption{Disk IMF for individual objects.}
\label{IMFdisk}}
{\centerline{\psfig{file=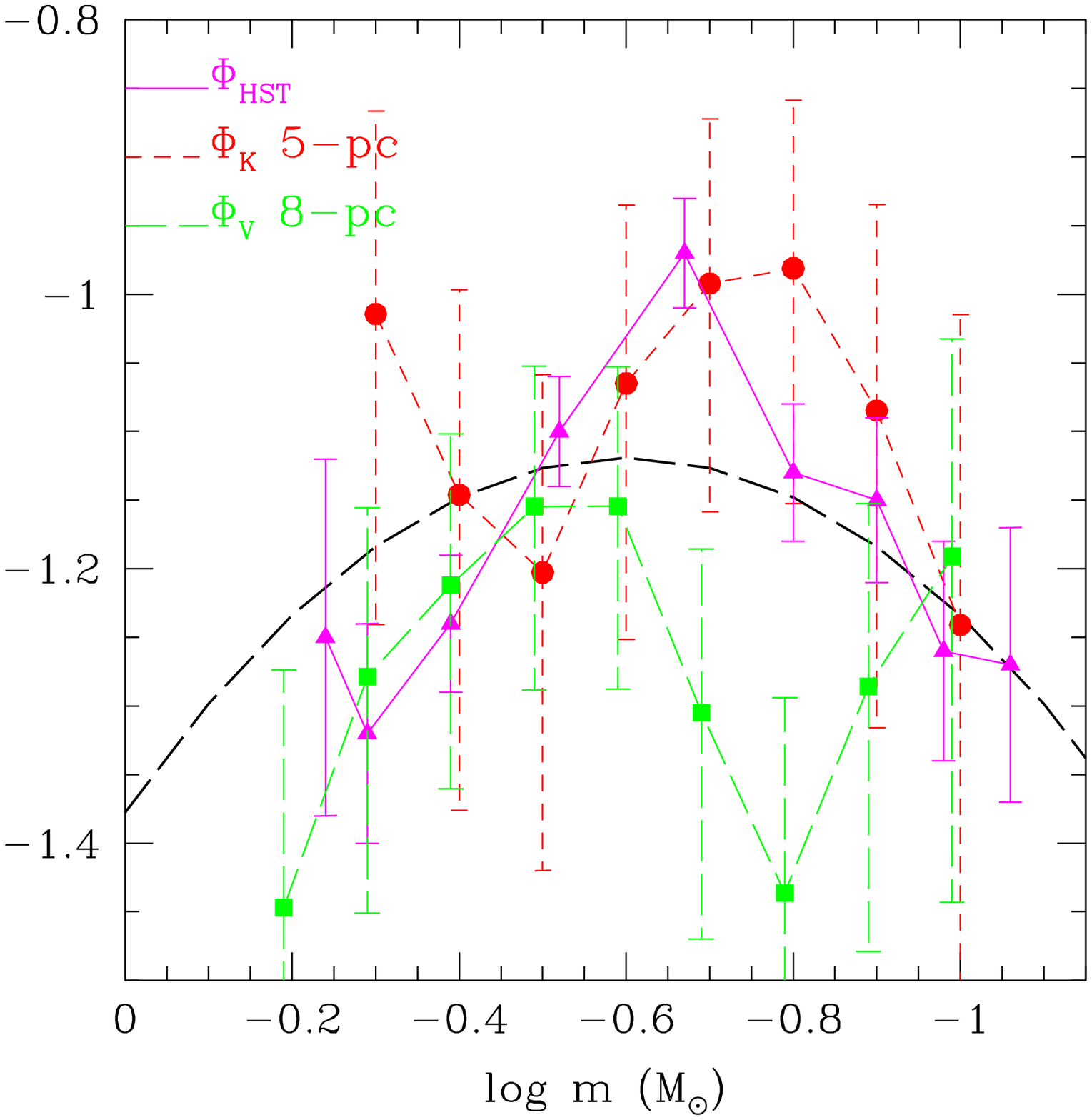,height=3in}}
\caption{Disk IMF for unresolved systems.}
\label{IMFsys}}
\end{figure}

The IMFs, $\xi(\log \,m)={dn\over d\log m}$, derived from the $\Phi_V$ and $\Phi_K$ LFs are portrayed in Figure \ref{IMFdisk} below 1 $\msol$. 
Superposed to the determinations is the following analytical parametrization (in $\mlvol$):

\begin{eqnarray}
\xi(\log \,m)&=&0.093\times \exp\Bigl\{-{(\log \, m\,\,-\,\,\log \, 0.2)^2\over 2\times (0.55)^2}\Bigr\},\,\,\, m\le 1\,\msol \nonumber \\
&=&0.041\,m^{-1.35\pm 0.3}\,\,\,\,\,\,\,\,\,\,\,\,\,\,\,\,\,\,\,\,\,\,\,\,\,\,\,\,\,\,\,\,\,\,\,\,\,\,\,\,\,\,\,\,\,\,\,\,\,\,\,\, , m\ge 1\,\msol
\label{IMF1}
\end{eqnarray}

This IMF differs a bit from the one derived in \cite{Chabrier03a}
since it is based on the revised 8-pc $\Phi_V$.
The difference at the low-mass end between the two parametrizations reflects the present uncertainty at the faint end of the disk LF, near the H-burning limit (spectral types $\ga$ M5).
Note that the field IMF is also
representative of the bulge IMF (triangles), derived from the LF of \cite{Zoccali00}.

A fundamental advantage of the nearby LF is the identification of stellar companions. It is thus possible to merge the resolved objects into multiple systems, to calculate the magnitude of the systems and then derive the {\it system} IMF. The following parametrization, slightly different from the one derived in \cite{Chabrier03a} is displayed in Figure \ref{IMFsys} (normalyzed as eqn.(\ref{IMF1}) at 1 $\msol$, where all systems are resolved):

\begin{eqnarray}
\xi(\log \,m)=0.076\times \exp\Bigl\{-{(\log \, m\,\,-\,\,\log \, 0.25)^2\over 2\times 0.55^2}\Bigr\},\,\,\, m\le 1\,\msol
\label{IMFsys}
\end{eqnarray}

As shown by \cite{Chabrier03b} and seen on the figure, this system IMF is in excellent agreement with the MF derived from the revised HST phometric LF (\cite{Zheng01}, showing that the discrepancy between the MF derived from the nearby LF and the one derived from the HST stemmed primarily from unresolved companions in the HST field of view.

\begin{figure}[ht]
\sidebyside
{\centerline{\psfig{file=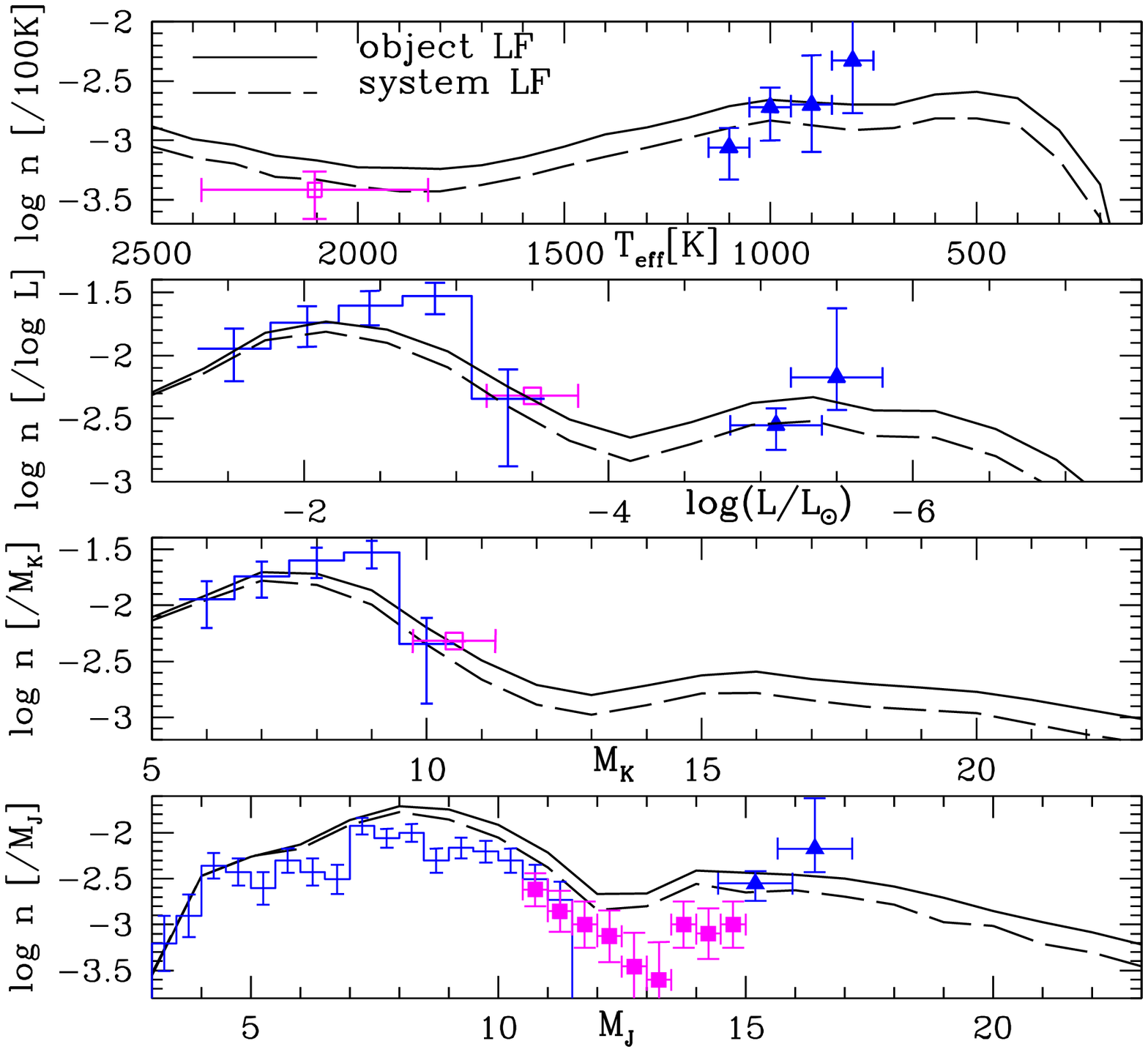,height=3.4in,width=3in}}
\caption{Comparison of the observed and theoretical LMS, L-dwarf 
($\blacksquare$) and T-dwarf ($\blacktriangle$) distributions.}
\label{BDLF}}
{\centerline{\psfig{file=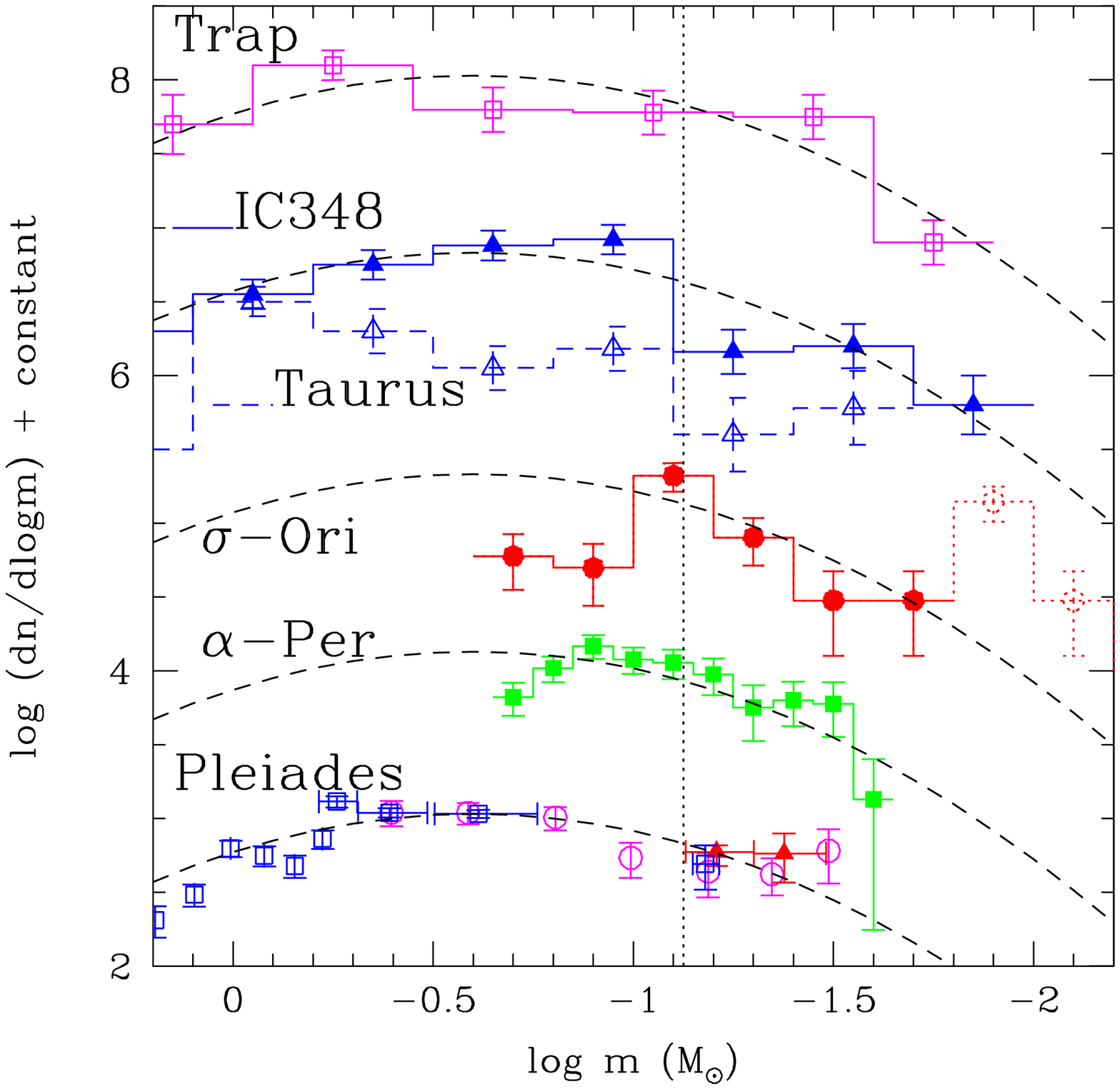,height=3.4in,width=2.6in}}
\caption{IMF for young clusters. Dash-line: field system IMF (2). Vertical dotted line: H-burning limit. }
\label{IMFclusters}}
\end{figure}

{\it {\bf The brown dwarf regime}}.

Many brown dwarfs (BD) have now been identified down to a few jupiter masses in the Galactic field with the DENIS, 2MASS and SDSS surveys. Since by definition BDs never
reach thermal equilibrium and keep fading with time, comparison of the predicted BD LFs,
based on a given IMF, with observations
requires to take into account time (i.e. formation rate) and mass (i.e. IMF) probability distributions. In the present review,
we proceed slightly differently from the calculations of \cite{Chabrier03a}. We start from the
{\it system} IMF (\ref{IMFsys}) and we include a probability distribution for the binary frequency
which decreases with mass. Indeed, various surveys now show that the
binary fraction $X_{bin}$ (and orbital separation) decreases with mass, varying
 from $X_{bin}\approx 60$\% for G and K-stars 
 to $\approx 40$\% for early (M0-M4) M-dwarfs,
to $\approx 20$\% for later M and L-dwarfs, correcting for undetected short period binaries, 
to $\approx 10$\% for T-dwarfs.

Figure \ref{BDLF}
displays the calculated BD density distributions as a function of $\te$, $L$, $\mk$ and $\mj$, based on the BD cooling models developed in the Lyon group, and the most recent estimated LMS and BD densities (\cite{Gizis00, Burg01,Cruz04}\footnote{Unpublished data in the J-band were kindly provided by N. Reid}).
The dash-line displays the distributions obtained with the system IMF (\ref{IMFsys}) while
the solid line corresponds to the IMF (\ref{IMF1}) for individual objects.
As mentioned above, the distributions obtained with this latter IMF are
consistent with a binary frequency decreasing from $\sim 50\%$ to $\sim 20\%$.
The agreement between the theoretical calculations and the observations is very satisfactory, keeping in mind the remaining uncertainties
in BD cooling theory and in accurate determinations of the observed BD $\te$, $\mbol$ and number densities. The predicted dip around $\mj\sim\mk\sim 13$, $\mbol\sim 15$ (\cite{Chabrier03a}) is confirmed by the recent L-dwarf observations.

\subsection{The young cluster mass function}

Figure \ref{IMFclusters} displays the MFs derived from the observed LFs of several young clusters, with ages ranging from $\sim$ 5 Myr to $\sim 150$ Myr, down to the substellar domain (see references in \cite{Chabrier03a}). Note that some of the faintest objects in $\sigma$-Or have been shown recently being field star contamination (\cite{McGovern04, Burg04}). We used the MMRs from BCAH
for the appropriate age in the appropriate observational filters. Accuracy of the BCAH
models for young clusters has been examined carefully by \cite{Luhman03}.
These observational surveys do not resolve multiple systems, so the derived MFs reflect the system MFs. Superposed to these IMFs is the field system IMF (\ref{IMFsys}). Figure \ref{IMFclusters} clearly points to a similar underlying IMF between young clusters and the Galactic field, except for the significantly less dense Taurus cluster.

\section{The globular cluster and spheroid mass function}

\begin{figure}[ht]
\sidebyside
{\centerline{\psfig{file=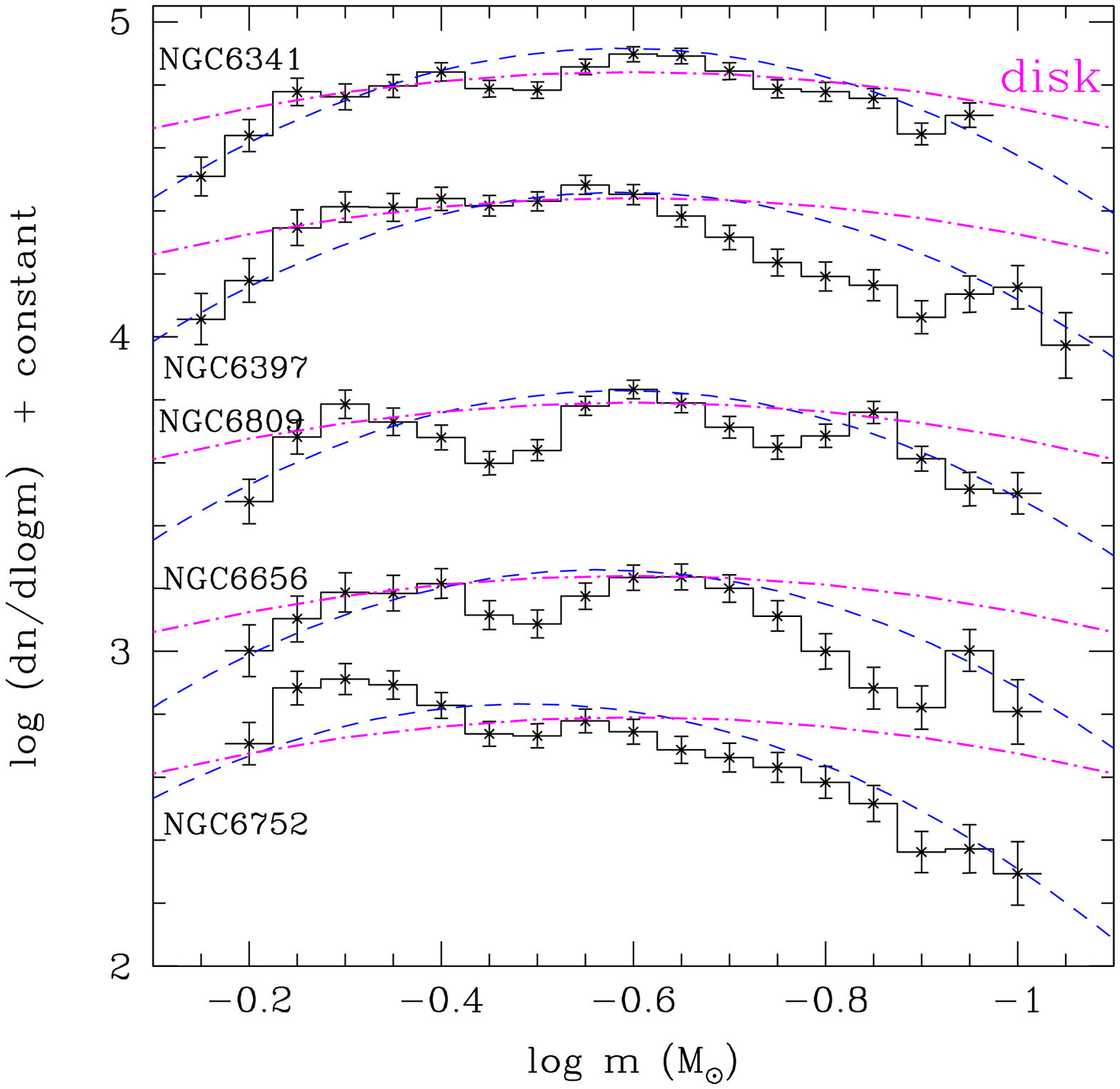,height=2.9in}}
\caption{IMF for several globular clusters. Dot-dash line: disk system IMF (2)}
\label{IMFGC}}
{\centerline{\psfig{file=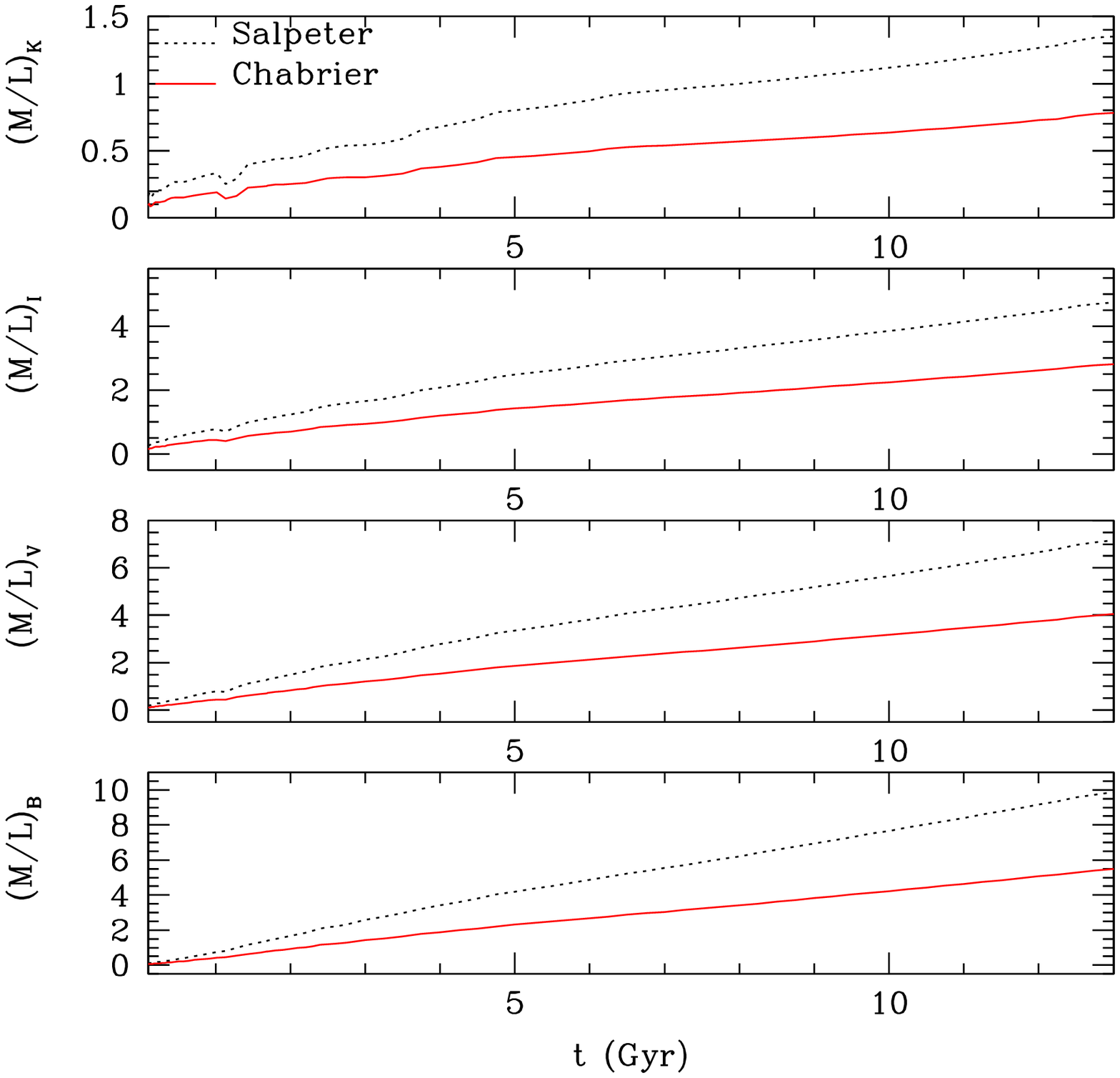,height=2.9in}}
\caption{Mass-to-light ratio for single stellar population in optical and NIR bands.}
\label{ML}}
\end{figure}

Globular clusters provide a particularly interesting test-bed to investigate the stellar MF. They provide a homogeneous
sample of MS stars with the same age, chemical composition and reddening, their distance is relatively
well determined, allowing straightforward determinations of the stellar LF.
From the theoretical point of view, the \cite{BCAH97} evolutionary models accurately reproduce the observed
color-magnitude diagrams of various clusters with metallicity $\mh \le -1.0$ both in optical and infrared colors, down
to the bottom of the main sequence,
with the limitations in the optical mentioned in \S2 for more metal-rich clusters. As mentioned above, however, the consequences of
this shortcoming on the determination of the MF remain modest. The IMFs derived by \cite{Chabrier03a} for several globular clusters, from the LFs observed with the HST, corrected for dynamical evolution, by \cite{ParDeMarchi00}
are displayed in Fig.\ref{IMFGC}. Superposed to the derived IMFs is the spheroid IMF given by eqn.(20) of \cite{Chabrier03a} (short-dash line), with the characteristic mass shifted by 1 to 2 $\sigma$'s ($m_c=0.22$ to $0.32$ $\msol$),
similar to the IMF derived by \cite{ParDeMarchi00}.
The disk IMF (2) is also superposed for comparison. We note the relative similarity of the these two IMFs, but a defficiency of very-low-mass objects, including BDs, in the cluster IMFs.

\section{Galactic implications : mass budget and mass to light ratio}

Integrating the IMF (\ref{IMF1}) yields the stellar and brown dwarf number- and mass-densities given in Table I, and the relative contributions $\mathcal{N}=N(\Delta m)/N_{tot}$ and
$\mathcal{M}=M(\Delta m)/M_{tot}$, where $N$ and $M$ denote respectively the number and mass of objects in the mass range $\Delta m$. Adding up remnant densities (\cite{Chabrier03a}) yields the stellar+substellar contributions to the Galactic mass budget. These new determinations give a BD-to-star number ratio of $n_{BD}/n_\star\sim 1/3$.

Figure \ref{ML} compares the M/L ratio obtained with the present IMF (\ref{IMF1}) and with the Salpeter IMF (see also \cite{BruzualCharlot03}). The present IMF yields M/L ratios in good agreement
with observations and with expectations from CDM hierarchical simulations of spiral disk galaxies or from dynamical
arguments (e.g. from Tully-Ficher relation), whereas a Salpeter IMF yields too large values (\cite{Portinari04}). 

\begin{table}
\caption[]{Present day stellar and brown dwarf Galactic budget$^{a,b}$.}
\bigskip
\begin{tabular*}{\textwidth}{@{\extracolsep{\fill}}lccc}
\sphline
\it Parameter&\it   Disk   & Spheroid &   Dark halo  \\
\sphline
\mbox{}\hspace{0.2cm} $n_{BD}$  & $2.6\times 10^{-2}$ & $ 3.5\times 10^{-5}$ & \\
\mbox{}\hspace{0.2cm} $\rho_{BD}$ & $1.0\times 10^{-3}$ &  $\la 2.3\times 10^{-6}$ & \\
\mbox{}\hspace{0.2cm} $n_{\star}$ & $(9.3\pm 2)\times 10^{-2}$ &  $\le (2.4\pm 0.1)\times 10^{-4}$ & \\
\mbox{}\hspace{0.2cm} $\rho_{\star}$ & $(3.4\pm 0.3)\times 10^{-2}$  & $\le (6.6\pm 0.7)\times 10^{-5}$ & $\ll 10^{-5}$ \\
\mbox{}\hspace{0.2cm} $n_{rem}$ & ($0.7\pm 0.1)\times 10^{-2}$  & $\le (2.7\pm 1.2)\times 10^{-5}$  &  \\
\mbox{}\hspace{0.2cm} $\rho_{{rem}}$ & ($0.6\pm 0.1)\times 10^{-2}$  & $\le (1.8\pm 0.8)\times 10^{-5}$ & $<10^{-4}$ \\
\mbox{}\hspace{0.2cm} $n_{{tot}}$ & $0.13\pm 0.03$ &  $\le 3.0\times 10^{-4}$ & \\
\mbox{}\hspace{0.2cm} $\rho_{{tot}}$ &($4.1\pm 0.3)\times 10^{-2}$ &   $\le (9.4\pm 1.0)\times 10^{-5}$ & $<10^{-4}$ \\
\hline \
\mbox{}\hspace{0.2cm} BD:\hspace{1.9cm} $\mathcal{N}$;  $\mathcal{M}$  & 0.20; 0.02  &
 0.10; 0.03 &  \\
\mbox{}\hspace{0.2cm} LMS($\le 1\msol$):\hspace{0.3cm} $\mathcal{N}$;  $\mathcal{M}$   & 0.71; 0.68   & 0.80; 0.77 &  \\
\mbox{}\hspace{0.2cm} IMS(1-$9\msol$):\hspace{0.5cm} $\mathcal{N}$;  $\mathcal{M}$  &0.03; 0.15 &  0.; 0. &  \\
\mbox{}\hspace{0.2cm} WD+NS:\hspace{1.1cm} $\mathcal{N}$; $\mathcal{M}$ & 0.06; 0.15 &
0.10; 0.20   &  \\
\sphline
\end{tabular*}
\begin{tablenotes}
$^a$The number densities $n$ are in [pc$^{-3}$], the mass
densities $\rho$ are in [$\mvol$].
\end{tablenotes}
\label{table.budget}
\end{table}

\section{Star formation theory}

\begin{figure}[ht]
\sidebyside
{\centerline{\psfig{file=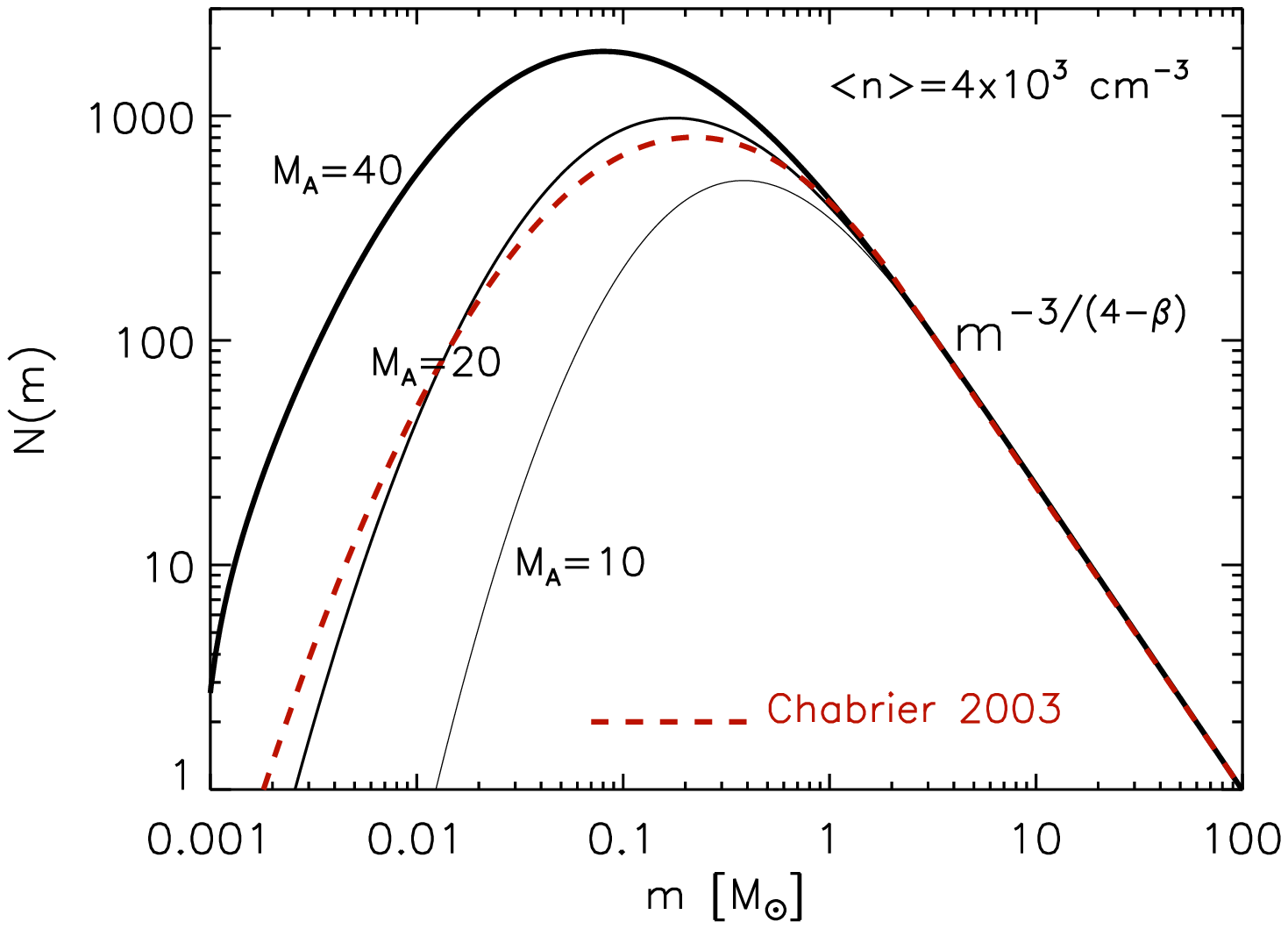,height=2.8in,width=2.78in}}
\caption{Comparison of the system IMF (2) with the one obtained by \cite{PadoanNordlund02}}}
{\centerline{\psfig{file=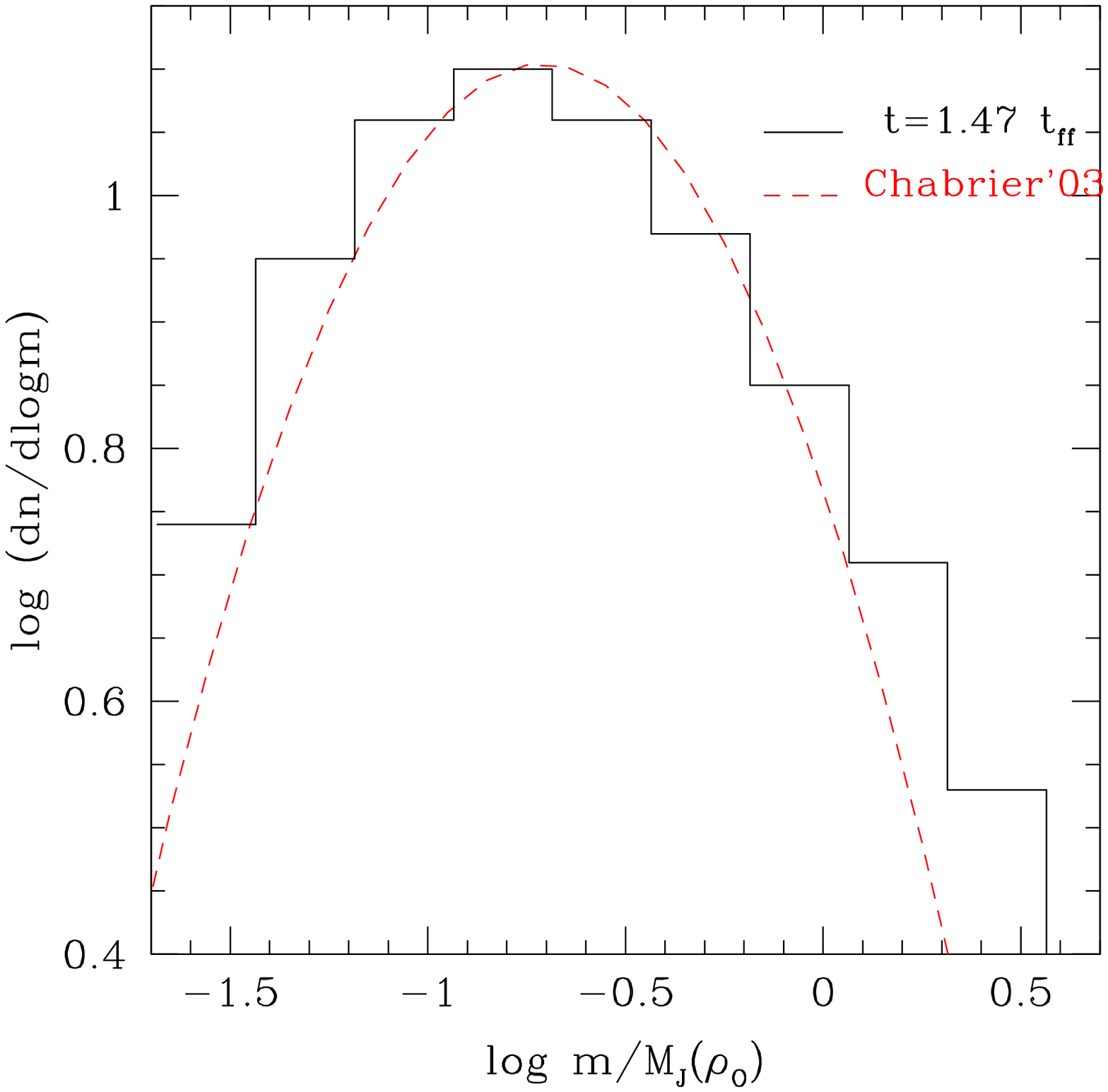,height=2.8in}}
\caption{Same with the one obtained from MHD simulations by \cite{Lietal04}.}}
\label{fig_MHD}
\end{figure}

Although we are still far from a general paradigm for star formation, some general
properties can be considered as robust : (1) star formation extends well below the H-burning limit, (2) the shape of the IMF seems to be very similar in very diverse environments, (3) star formation is a rapid process, comparable to the dynamical timescale $\tau_{dyn}=(3\pi/ 32G\rho)^{1/2}  \approx$1-5$\times 10^5$ yr for typical star-forming molecular clouds, (4) the stellar IMF seems to be reminiscent of the prestellar
core mass spectrum,
suggesting that
the IMF is already determined by the cloud clump mass distribution (see Andr\'e and Myers, this conference), (5) on large scales, the spectral line widths in molecular clouds indicate supersonic, super-Alfv\'enic conditions.
All these observations point to a common driving mechanism for star formation, namely
turbulence-driven fragmentation. Figure \ref{fig_MHD} compares the mass spectrum obtained
from MHD simulations done
recently by \cite{PadoanNordlund02} and \cite{Lietal04} with the {\it system} IMF (\ref{IMFsys}) (indeed, the prestellar cores correspond to multiple systems, which eventually will fragment further into individual objects). Although such comparisons must be considered with
due caution before drawing any conclusion, the agreement between the simulations and the
IMF representative of the field is amazing. Note that such hydrodynamical simulations form BDs in
adequat numbers from the same fragmentation mechanism as for star formation. Various
observations of disk accreting BDs indeed show that BDs and stars form from the same underlying mechanism. Motivation for invoking the formation of BDs by ejection thus no
longer seems to be necessary. Moreover, the recent observation of a wide binary BD (Luhman 2004) definitely contradicts the predictions of such a scenario as a dominant formation mechanism for BDs. This new picture thus combines turbulence, as the initial driving mechanism for fragmentation, and gravity, providing a natural explanation for a (scale free)
power-law IMF above a critical mass, namely the mean thermal Jeans mass $\langle m_J \rangle$, and a lognormal distribution below, due to the fact that only the densest cores, exceeding the local Jeans mass, will collapse into bound objects.
Note that in these simulations of supersonic turbulence, 
only a few percents of the total mass end up into the collapsing cores
after one dynamical time, solving naturally
the old high efficiency problem associated with turbulence-driven star formation.
Recent similar simulations by \cite{LiNakamura04}, on the other hand, suggest
enhanced ambipolar diffusion to occur through shock compression.

\section{Conclusion and perspective}

In this review, we have examined the most recent determinations of the Galactic stellar IMF.
Thanks to tremendous progress both in observational techniques and in the theory of low-mass stars and brown dwarfs, the IMF can now be determined down to a few jupiter masses, two orders of magnitude below the $\sim 0.5\msol$ limit of accuracy
of the Salpeter (1955) IMF.
This IMF adequately reproduces various observational constraints, star and BD
counts, binary frequencies, galactic mass-to-light ratios. It is well described by a Salpeter power-law above $\sim 1\msol$ rolling down into a lognormal form below this mass, with a characteristic mass around 0.2 $\msol$, although more data is needed to really nail down this issue. Only for
nearly zero-metal environments in the early universe do we expect a significantly larger characteristic
mass, due to the lack of efficient cooling mechanism in the cloud (see Bromm, this conference).
The universality of the IMF in various environments, disk, young and globular clusters, spheroid, points to
a universal triggering mechanism and a dominant cooling process for star formation.
Small scale dissipation of large scale supersonic MHD turbulence provides an appealing 
solution for this mechanism.


\end{document}